\documentclass[aps,preprint]{revtex4-1}

\usepackage{natbib}
\usepackage{graphicx}

\usepackage{color}
\usepackage{commath}
\usepackage{graphicx}
\usepackage{bm}
\usepackage{verbatim}

\begin{document}

\title{Distinct magnetic field dependence of N\'eel skyrmion sizes in ultrathin nanodots}
\author{F. Tejo$^1$}
\author{A. Riveros$^1$}
\author{J. Escrig$^{1,2}$}
\author{K. Y. Guslienko$^{3,4}$}
\author{O. Chubykalo-Fesenko$^5$} 

\affiliation{$^{1}${ Departamento de F\'{\i}sica, Universidad de Santiago de Chile (USACH), 9170124 Santiago, Chile }\\
$^{2}${ Center for the Development of Nanoscience and Nanotechnology (CEDENNA), 9170124 Santiago, Chile}\\
$^{3}$Departamento de F\'{\i}sica de Materiales, Universidad del Pa\'{\i}s Vasco, UPV/EHU, 20018 San Sebastian, Spain\\
$^{4}$IKERBASQUE, the Basque Foundation for Science, 48013 Bilbao, Spain\\
$^{5}$Instituto de Ciencias de Materiales de Madrid, 28049 Madrid, {Espa\~{n}a}}

\begin{abstract}
We investigate the dependence of the N\'eel skyrmion size and stability on perpendicular magnetic field  in ultrathin circular magnetic dots with out-of-plane anisotropy and interfacial Dzyaloshinskii-Moriya exchange interaction. Our results show the existence of two distinct dependencies of the skyrmion radius on the applied field and dot size. In the case of skyrmions stable at zero field, their radius strongly increases with the field applied parallel to the skyrmion core  until skyrmion reaches the metastability region and this dependence slows down.  More common metastable skyrmions demonstrate a weaker increase of their size as a function of the field  until some critical field value at which these skyrmions drastically increase in size showing a hysteretic behavior with coexistence of small and large radius skyrmions and small energy barriers between them.
The first case is also characterized by a strong dependence of the skyrmion radius on the dot diameter, while in the second case this dependence is very weak.

\end{abstract}
\maketitle

\section{Introduction}

Magnetic skyrmions attracted recently enormous attention of researchers due to their promising static and dynamical  properties resulting from their chiral structure and non-trivial topology \cite{Fert,Nagaosa,Fert2017}.
 Here we consider isolated N\'eel-type magnetic skyrmions stabilized  in ultrathin circular dots with Dzyaloshinskii-Moriya interactions  (DMI) coming from the interfaces between transition metals (Co) and heavy metals with large spin-orbit coupling (Pt, Ir) \cite{Yang, Moreau2016}. The choice of these skyrmions is based on their promising technological perspectives.
First, several articles report that the N\'eel skyrmions  can be stable at room temperature opening the possibility for their use in information technology \cite{Moreau2016,Boulle2016, Woo}.
Secondly,  skyrmions promise high thermal stability. This hypothesis comes from the idea of their topological protection,  although no rigorous proof of the high thermal stability based on the energy barrier calculations exist \cite{Rohart2,Bessarab}.
 Several articles also report on the possibility to move the N\'eel skyrmions by spin-polarised currents, especially with small current density values \cite{Woo,Fert,Sampaio2013} and to detect them by the spin-Hall effect \cite{Jiang,Litzius}. All these properties are extremely useful for future spintronic applications. These findings make skyrmions candidates for low energy consumption applications and several designs for the  skyrmions-based spin-torque nano-oscillators, microwave detectors and logic devices were recently introduced \cite{Tomasello,Yu,Zhou,Fert2017}.

The skyrmion stability and their equilibrium properties are important issues for any device applications. The skyrmion stability depends on all micromagnetic parameters, including the DMI, magnetic anisotropy strength, the  exchange stiffness, and the applied magnetic field.
 Rohart and Thiaville \cite{Rohart2013} have suggested a simple formula coming from the cylindrical domain wall theory for the critical DMI strength $D$ at which the skyrmion becomes stable, $D > D_c=(4/\pi)\sqrt{AK_{\text{eff}}}$, where $A$ is the exchange stiffness parameter, $K_{eff}=K_{u}- \mu_{0}M_{s}^2/2$ is the  effective magnetic uniaxial anisotropy constant which includes the magnetostatic contribution, and $M_{s}$ is the material saturation magnetization.
In spite of the fact that this approach is not well justified \cite{Beg2015,Vidal2017}, it gives a simple and frequently working estimation for the skyrmion stability edge.  Importantly, most of the N\'eel skyrmions are metastable states in absense of the applied magnetic field \cite{Aranda}, i.e., have the magnetic energy larger than the perpendicularly magnetized states.

In many experimental works several skyrmions are uncontrollably present. For applications, however,  individual skyrmions should be nucleated and manipulated. This can be done by geometrical confinement in patterned magnetic films (stripes, dots, etc.) with out-of-plane magnetic anisotropy.
Recent calculations show that the skyrmion stability region is significantly modified due to the presence of magnetic sample boundaries \cite{Guslienko} and essential contribution of the magnetostatic energy.
It has been demonstrated that the skyrmion stability depends on the dot size and the stability region is larger in dots with smaller sizes \cite{Aranda}.

Following the approach of Rohart and Thiaville \cite{Rohart2013}, the radius of skyrmions has been predicted to reveal a strong increase as a function of the DMI parameter when approaching the critical value $D_c$  for the metastability region.
As a function of the applied magnetic field, the skyrmion radius (when the skyrmion core magnetization direction is antiparallel to that of the field) decreases as reported experimentally in Ref. \cite{Moreau2016} and increases when the core direction is parallel to the field direction, see also Ref. \cite{GarciaSanchez}.

In this article we systematically investigate the dependence of the Neel skyrmion diameter on applied out-of-plane magnetic field in ultrathin circular magnetic dots with different sets of the micromagnetic parameters and dot sizes. We found a very distinct behavior of the skyrmion radius depending on the fact whether the skyrmion configuration is stable at zero field or metastable.

\section{Model}
\begin{figure}[hbtp]
\centering
\includegraphics[width=7cm]{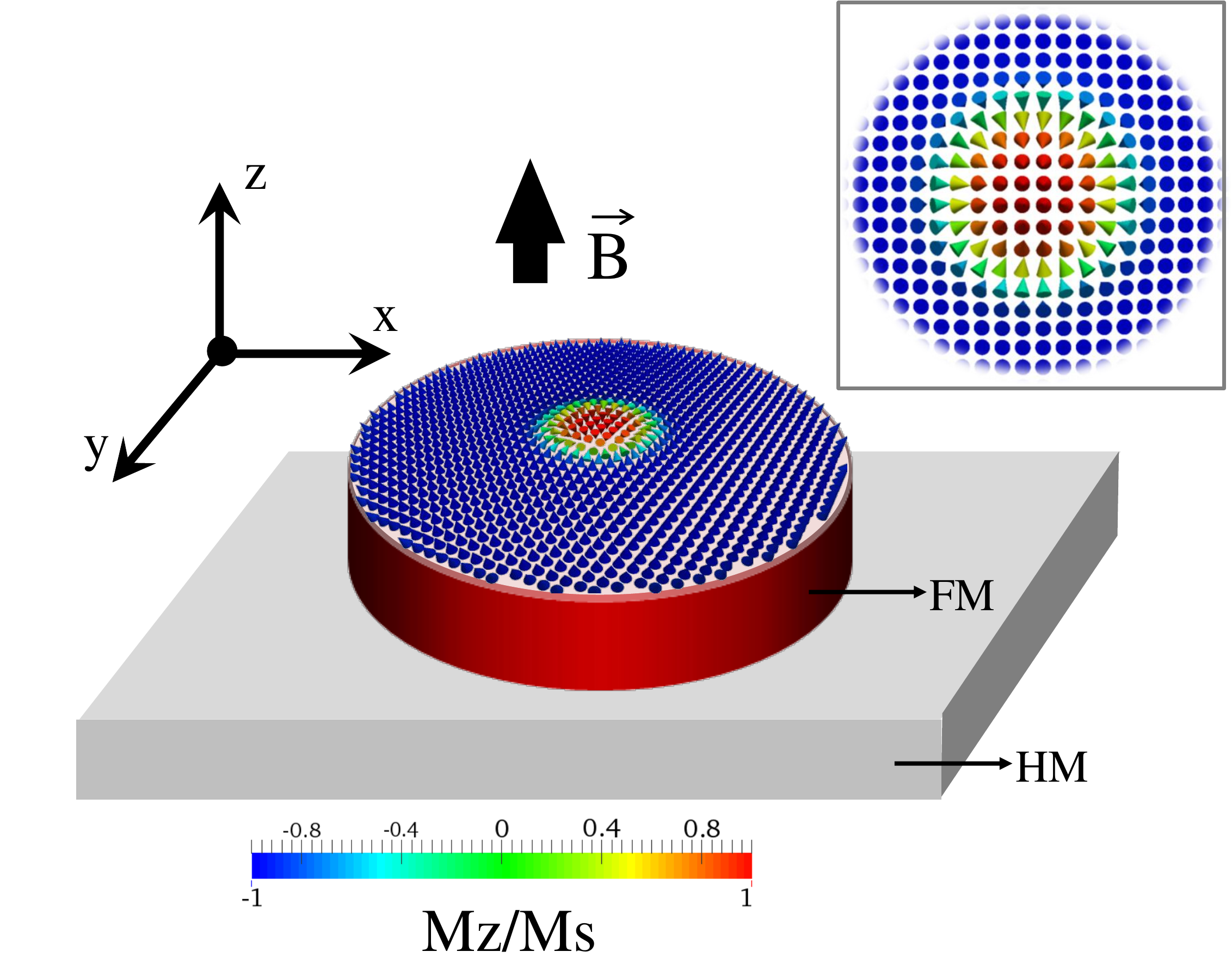}
\caption{ Ultrathin circular ferromagnetic (FM) nanodot on a heavy metal (HM) substrate. The applied magnetic field ($\vec{B}$) is in the $z$-direction and is perpendicular to the dot plane.}
\label{fig:Fig1}
\end{figure}

To study the dependence of the skyrmion diameter and its stability in a ultrathin circular magnetic dot varying an out-of-plane external magnetic field we apply the micromagnetic approach. The geometry of physical system under study is depicted in Fig. \ref{fig:Fig1}, where we indicate that in our convention the positive out-of-plane magnetic field is parallel to the skyrmion core, while the negative field is applied anti-parallel to it.  The dot has variable diameter $d$ and fixed thickness of 0.6 nm. We consider  different materials of the layers with the parameters taken from the literature and corresponding to different CoPt-based multilayer systems which we schematically indicate as  PtCoPt\cite{Sampaio2013}, PtCoMgO\cite{Boulle2016}, IrCoPt\cite{Moreau2016} and PtCoNiCo\cite{Ryu2014}.

The system energy functional is defined by the following energy density
\begin{eqnarray}
\label{Edensity}
\nonumber &\varepsilon(\vec{m})&= A\sum_{\alpha=x,y,z}(\vec{\nabla} m_\alpha)^2+D[m_z \vec{\nabla}\cdot\vec{m}-(\vec{m}\cdot\vec{\nabla})m_{z}]-\\
&K_u& m_z^2 -\dfrac{M_s}{2}\mu_{0}\vec{m}\cdot\vec{H}_d-M_s\vec{m}\cdot\vec{B},
\end{eqnarray}
where the first term corresponds to the exchange energy with the stiffness constant $A$, the second term stands for the DMI with the  constant $D$, the other terms correspond to the out-of-plane uniaxial anisotropy (where $K_u$ is the uniaxial anisotropy constant), the magnetostatic  and the Zeeman energies, respectively. The dot micromagnetic parameters are summarised in Table I. The material quality parameter $Q>1$ for each set of the parameters and $D$ is large enough to ensure the Neel skyrmion metastability/stability in the zero out-of-plane magnetic field.

For the micromagnetic simulations we used the object-oriented micromagnetic framework (OOMMF) code with the extension accounting for the interfacial DMI\cite{Rohart2013}.  The dot volume was discretized in cubic cells with the cell sizes of 0.5$\times$0.5$\times$0.6 nm$^3$. The out-of-plane magnetic field was varied in the interval from -20 mT to 30 mT.


To better understand the dependence of skyrmion diameter as a function of magnetic field we also used a  semi-analytical approach. For this we defined the reduced magnetization vector $\vec{m}$ of the skyrmion via the spherical angles $(\Theta, \Phi)$. The polar angle $\Theta = \Theta_{0}(\rho)$ is a circularly symmetric function  of the polar coordinate $\vec{\rho} =(\rho, \phi)$ and the azimuthal angle is $\Phi= \Phi_{0} + \phi$, where $ \Phi_{0} = 0, \pi$ for the case of the Neel (hedgehog) skyrmion. For the description of the skyrmion configuration we used the following trial function

\begin{eqnarray}
\tan{\dfrac{\Theta_{0}(r) }{2}=\dfrac{r_{s}}{r}}e^{\xi(r_{s}-r)},
\label{Anzats}
\end{eqnarray}
where  $r_{s} = R_s/l_{ex}$ is the reduced skyrmion radius (expressed through the exchange length $l_{ex}=\sqrt{2A/\mu_{0}M^2_s}$  and $r = \rho/l_{ex}$, $\xi^2=Q-1$ and $Q=2K_u/\mu_{0}M^2_s$ is the quality factor. The trial function given in Eq.(\ref{Anzats}) was previously used in Ref. [22] to describe the axially symmetric magnetic solitons in a 2D Heisenberg easy axis ferromagnet showing a good agreement with the direct energy minimization method. More recently it has been also successfully used to calculate the skyrmion stability diagram at zero field and the dependence of the skyrmion radius on the DMI strength and temperature \cite{Tomasello2} also showing a good agreement with micromagnetic modelling. For the isotropic case ($ \xi=0$) Eq. (\ref{Anzats}) recovers the Belavin-Polyakov solution  and leads to the finite exchange energy at $r \rightarrow 0$.


The equilibrium skyrmion size is found by minimizing the energy functional $E = \int dV \varepsilon(\vec{m})$, using the energy density $\varepsilon(\vec{m})$ given in Eq. \eqref{Edensity}
\begin{eqnarray}
\frac{\partial{E}}{\partial{r_{s}}}=0,
\;\;\; \frac{\partial^2{E}}{\partial{r_{s}}^2}>0
\end{eqnarray}

\begin{widetext}
\begin{table}[]
\centering
\label{parameters}
\begin{tabular}{|lcccc|}

\multicolumn{5}{l}{}   \\  \hline

Parameter & PtCoPt\cite{Sampaio2013} & PtCoMgO\cite{Boulle2016} & IrCoPt\cite{Moreau2016} & PtCoNiCo\cite{Ryu2014} \\ \hline

Saturation magnetization Ms (10$^3$ A/m)   & 580     & 1400    & 956     & 600     \\
Exchange constant A (10$^{-12}$ J/m)  & 15     & 27.5       & 10     & 20   \\

Perpendicular anisotropy constant K$_{u}$ (10$^6$ J/m$^3$) & 0.7     & 1.45    & 0.717  & 0.6    \\

DMI parameter D (10$^{-3}$ J/m$^2$)  & 3  & 2.05    & 1.6      & 3     \\ \hline
\end{tabular}
\caption{Micromagnetic parameters used in the simulations.}
\end{table}
\end{widetext}

\section{Results and discussion}
\subsection{Micromagnetic simulations}

\begin{figure}[hbtp]
\centering
\includegraphics[width=7.5cm]{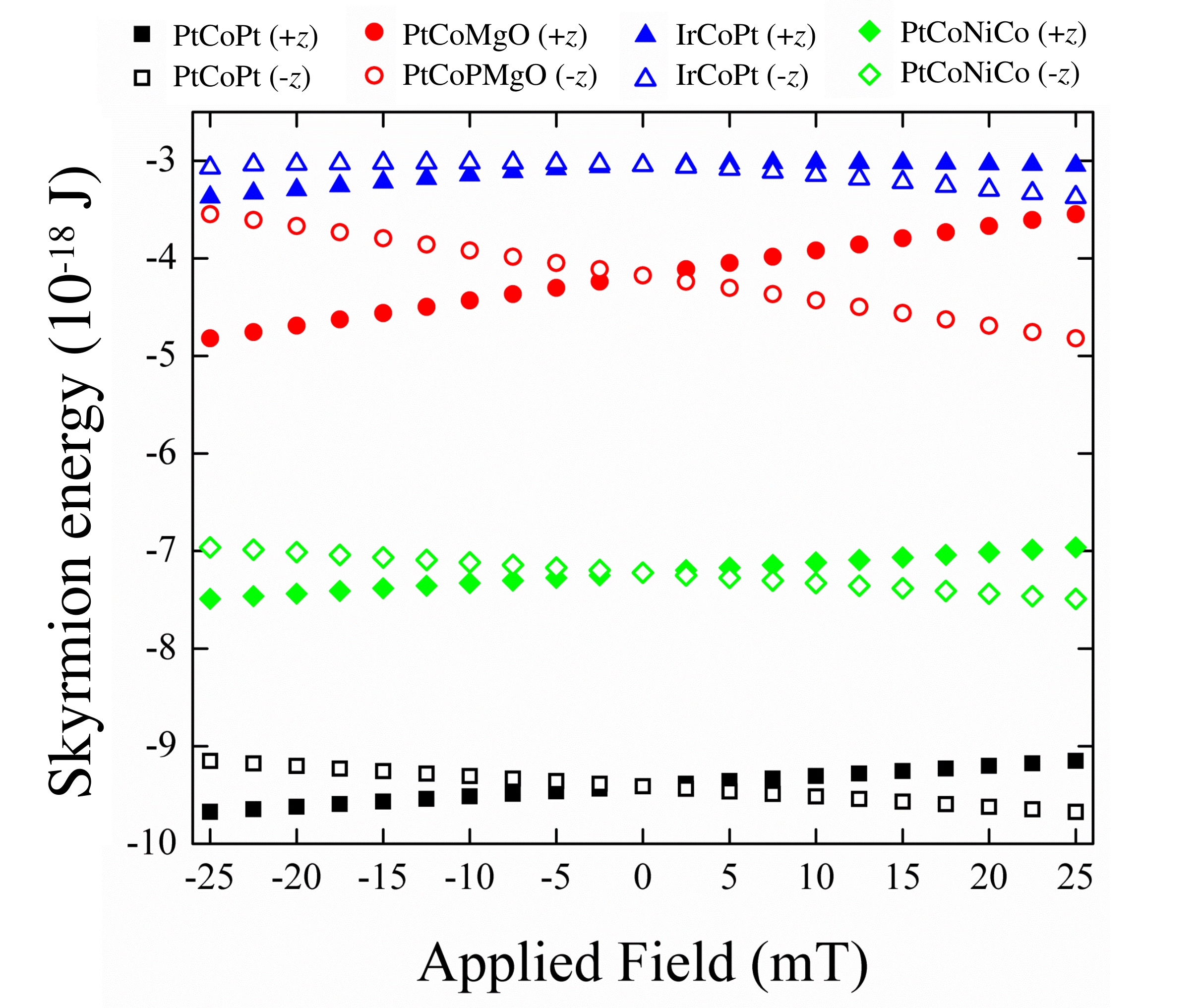}
\caption{Numerical results obtained for the skyrmion energy as a function of the out-of-plane magnetic field for different materials and orientations of the skyrmion core with respect to the field. The dot diameter is 200 nm.
}

\label{Coreupdown}
\end{figure}

Figure \ref{Coreupdown} shows the comparison between the magnetic energies of the skyrmions with the core parallel and antiparallel to the bias field directions. The results indicate that  skyrmion with the core antiparallel to the field direction always has lower energy than skyrmion with the core parallel to the field \cite{Riveros2017}. In any case, the results are symmetric with respect to the field inversion. However, we should remember that in our convention for the positive field there is always a skyrmion with a lower energy which can be obtained by symmetry from the results for negative fields.

\begin{figure}[hbtp]

\centering
\includegraphics[width=7cm]{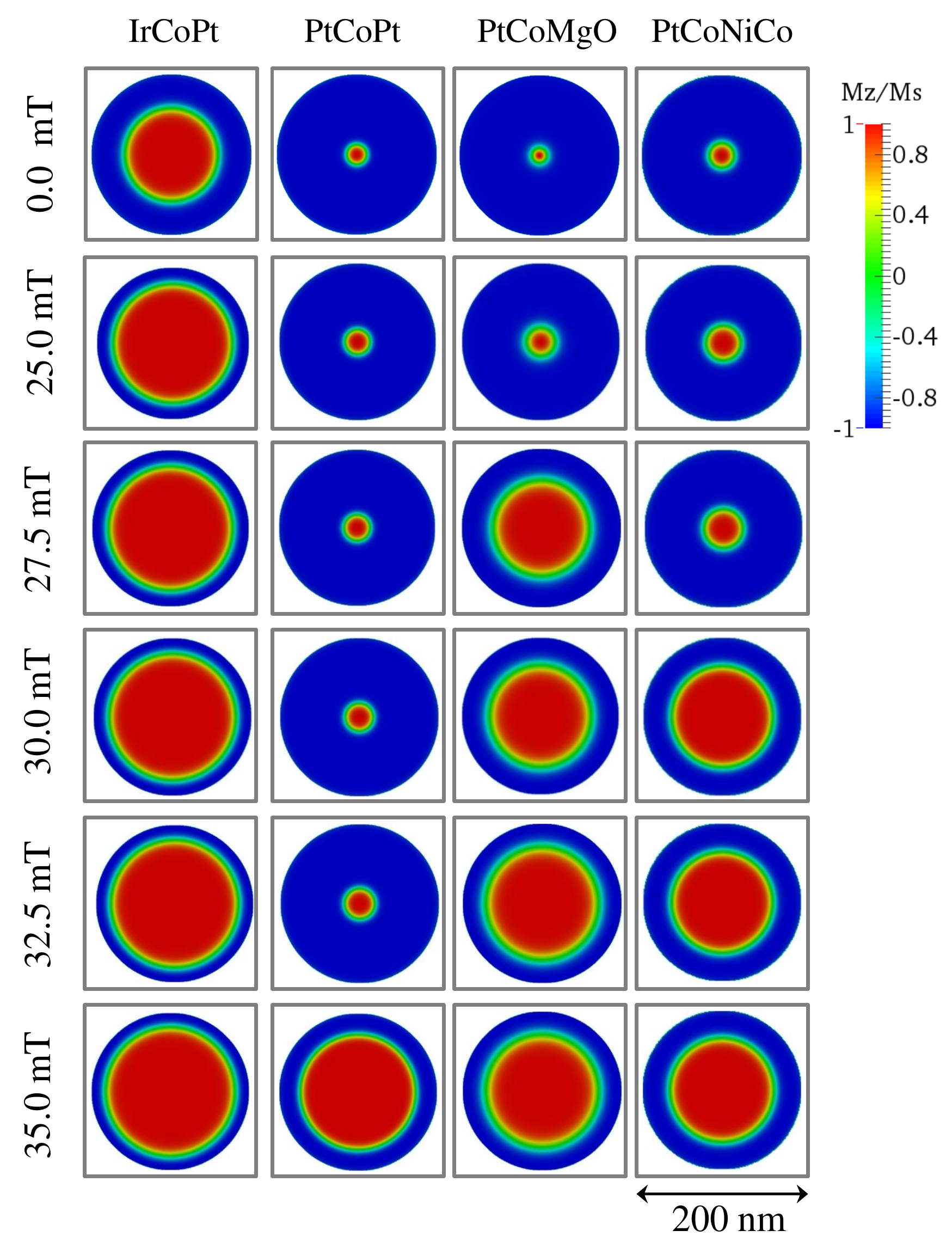}
\caption{Snapshots of the Neel skyrmion magnetization configurations in a cylindrical dot of 200 nm  diameter showing two types of skyrmions, a small one and a large one, stabilized by the dot boarder. Note that the initial configurations in all cases correspond to that of the skyrmion at zero out-of-plane magnetic field. }
\label{Snapshots}
\end{figure}

Figure \ref{Snapshots} illustrates the snapshots for the Neel skyrmion magnetization distribution in different materials for the circular dots with diameter $d$ = 200 nm at different applied field values. Two distinct situations can be observed when the value of the magnetic field increases. For IrCoPt, the skyrmion is large and occupies almost the whole dot. In all other cases, the skyrmion is small for certain magnetic field values, but at some magnetic field it drastically increases its size until it becomes also large.

To tackle the difference between these situations we present below  in Fig. \ref{compare} the comparison between the energies of the skyrmion state and the perpendicularly magnetized state applying the out-of-plane magnetic field in the positive (parallel to the skyrmion core) and negative (antiparallel to the skyrmion core) directions. For simplicity we present here the skyrmion with the core parallel to the $z$-axis. The other skyrmion energies can be easily seen by using the symmetry with negative fields.  Also for the perpendicularly magnetized case (quasi-uniform state) we only indicate that of the smallest energy, i.e., with the skyrmion core parallel to the field.
The important conclusion of the energy considerations is that in the case of IrCoPt the skyrmion is a ground state at zero field and there exists a region of its stability in some field interval around. In all other considered cases such as PtCoPt presented in Fig. \ref{compare} the skyrmion is  metastable. Note that only for one of the four considered parameter sets the skyrmion is the system ground state. This is in agreement with our recent micromagnetic simulations showing that the stable skyrmions are very rare \cite{Aranda} and most of the skyrmions reported in the literature are metastable.

\begin{figure}[hbtp]
\centering
\includegraphics[width=8cm]{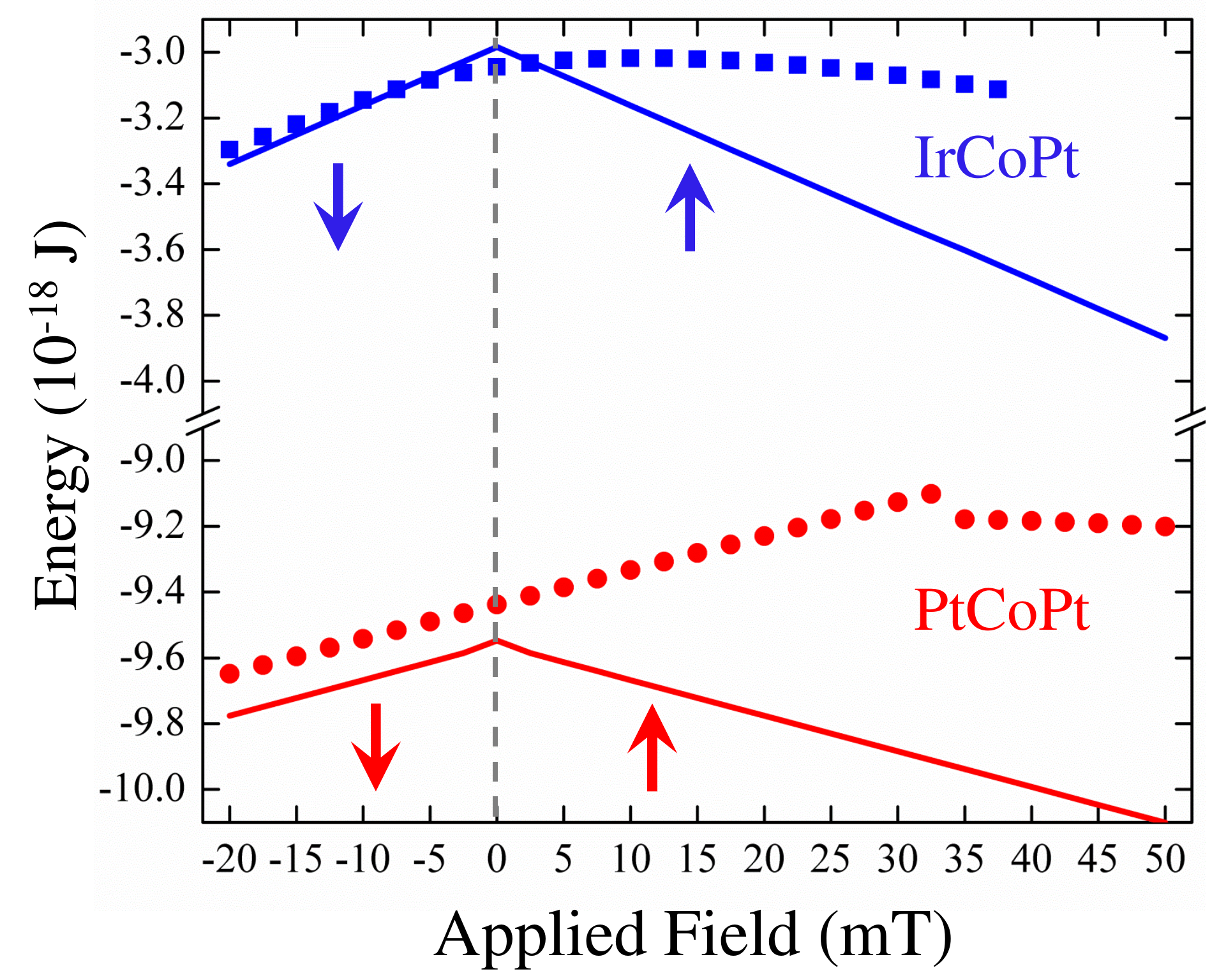}

\caption{Comparison of the energies between a ferromagnetic state and a skyrmion state for the   ultrathin nanodot of 200 nm in diameter made of IrCoPt (upper curves) and PtCoPt (lower curves) materials. The symbols correspond to the skyrmion state, while the solid lines corresspond to the ferromagnetic state,  which is always parallel to the out-plane magnetic field.}
\label{compare}
\end{figure}

\begin{figure}[hbtp]

\centering
\includegraphics[width=8cm]{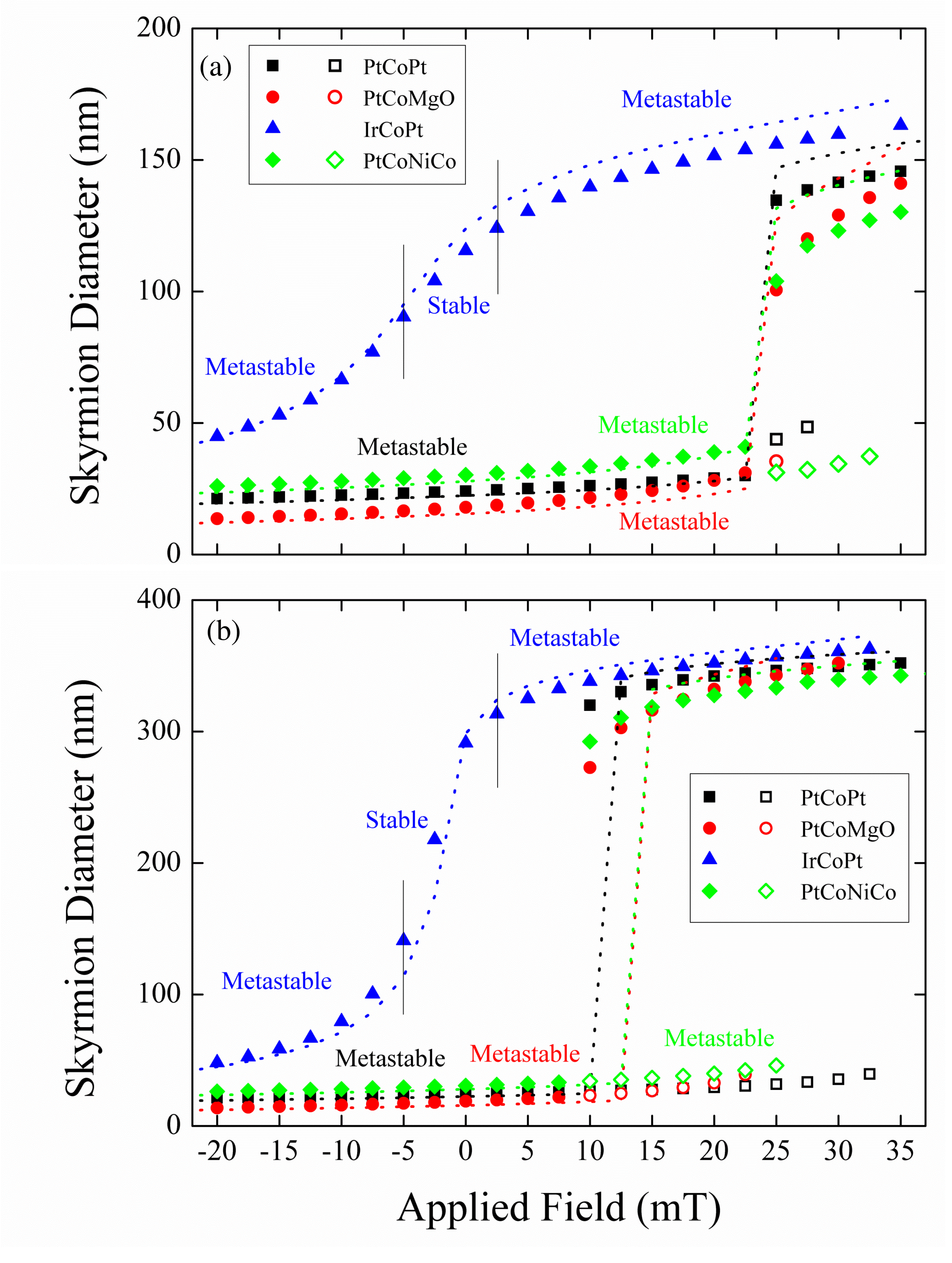}
\caption{Dependence of the skyrmion diameter on the applied out-of-plane magnetic field for a ultrathin cylindrical dot of diameter (a) $d = 200$ nm and (b) $d = 400$ nm. Open symbols represent  small radius skyrmion  states with higher energy than the skyrmion  states of larger diameter in the region of bi-stability. The dotted lines stand for the analytical calculations using the skyrmion profile given by Eq. (2). Here we plot the skyrmion sizes only for the skyrmion states of smaller energy.}
\label{Diameter}
\end{figure}

Our main results are presented in Fig. \ref{Diameter}, where we show the dependence of skyrmion radii on the applied field values for the dots with two diameters  $d = 200$ nm and $d = 400$ nm  and four chosen sets of the material parameters (see Table I). Note existence of two distinct characteristic dependencies. In the case of IrCoPt, the skyrmion is stable at zero field. Its diameter is large at zero field and strongly increases when the field increases until the skyrmion becomes metastable and the field dependence becomes weak.
In the other cases, the skyrmion is always metastable and its field dependence is weak until some critical field value is reached, where a sudden increase of the skyrmion diameter takes place. Around this field there are two types of the metastable skyrmions: a small and a large one and the field dependence of the skyrmion radius is characterized by a hysteretic behavior as a function of the applied field. In this region the skyrmion of the larger diameter (i.e., stabilized by the dot boarder) has a smaller energy than the small radius skyrmion. Note that the critical magnetic field for the transition between small and large size skyrmion is larger in the smaller dot. The hysteresis width is larger in larger dots.

\begin{figure}[hbtp]
\centering
\includegraphics[width=8.5cm]{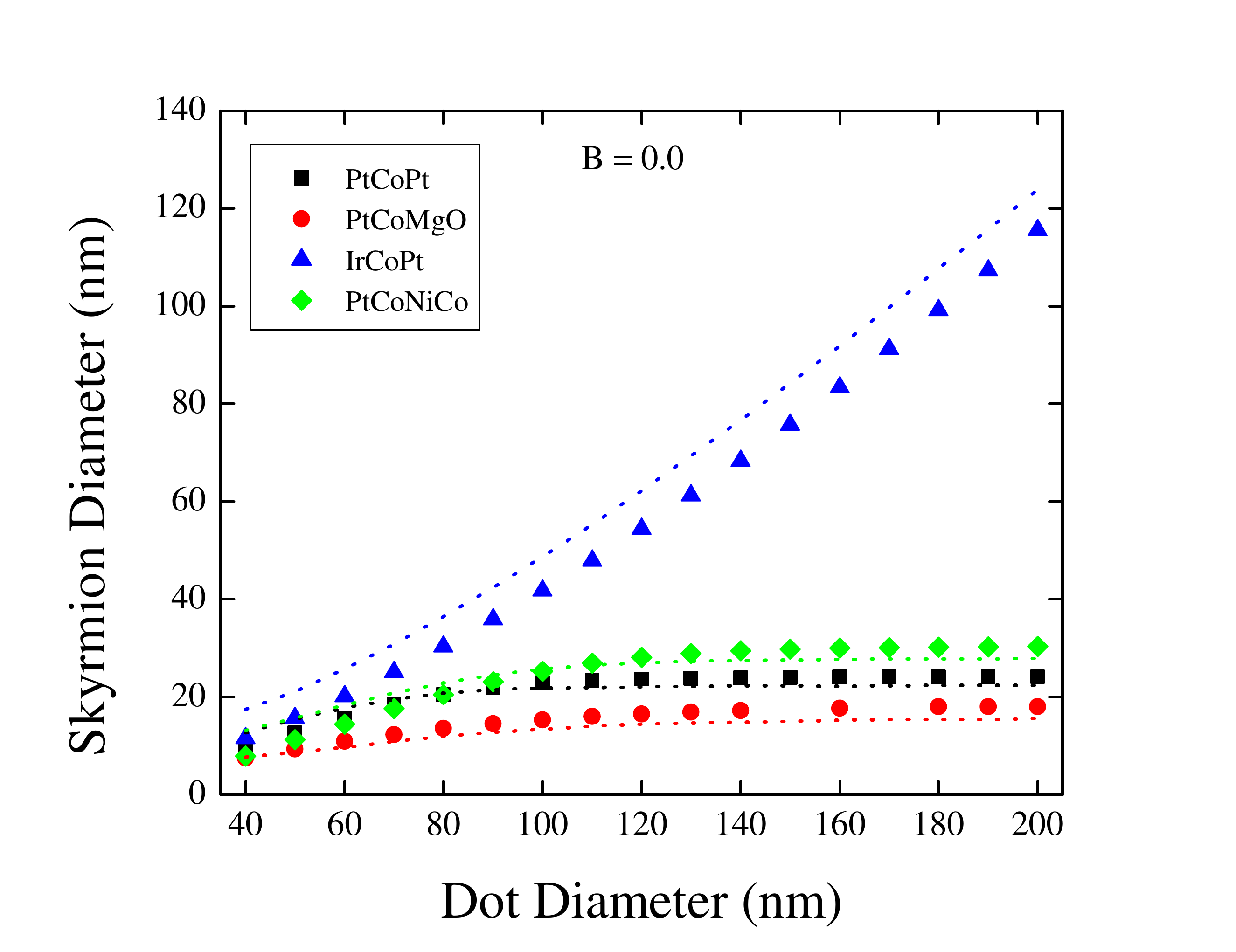}
\caption{Simulated dependencies of the skyrmion diameter on the nanodot diameter for different materials and applied magnetic field $B$=0. The dotted lines represent theoretical predictions.}
\label{DDiam}
\end{figure}

Figure \ref{DDiam} presents the skyrmion diameter as a function of the dot diameter at zero out-of-plane field. In very small dots all skyrmions have similar dimensions stabilized by the dot boarder.
While the small metastable skyrmions reveal a weak increase of their diameter  with the diameter of the dot, the large stable skyrmions reveal a strong dependence keeping the ratio between the diameter of the skyrmion and the dot approximately as constant.

\subsection{Analytical model}

\begin{figure}[hbtp]
\centering
\includegraphics[width=7.5cm]{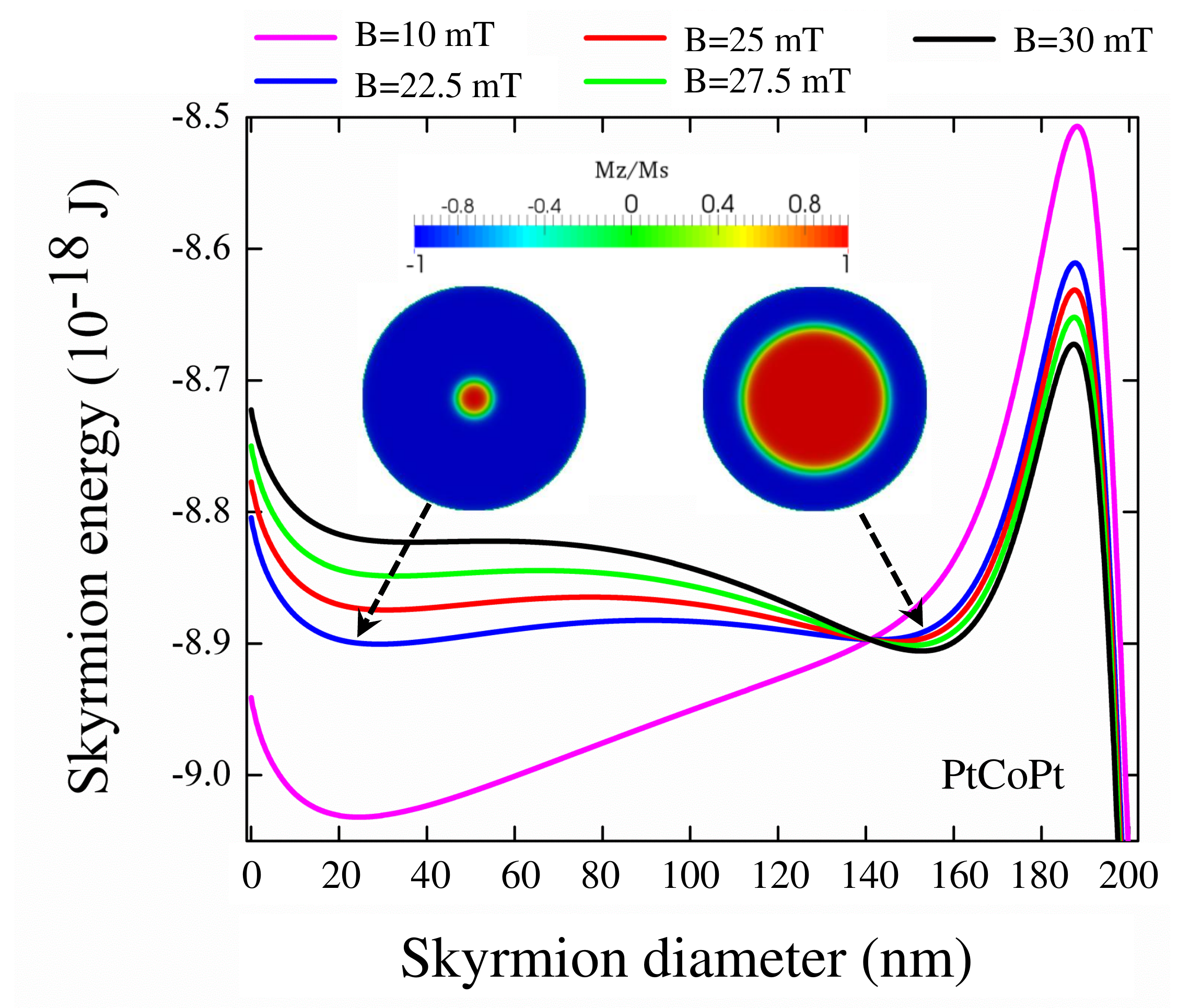}
\caption{Skyrmion energy in PtCoPt dot of 200 nm in diameter, as function of the skyrmion diameter for different values of the out-of-plane magnetic field near the critical value calculated by analytical approach.}
\label{Energy_D}
\end{figure}

The results of the analytical model  are presented by the dotted lines in Fig. \ref{Diameter} and \ref{DDiam} as a function of the out-of-plane field and as a function of the dot diameter, respectively. For simplicity the lines indicate the skyrmion of the lowest energy only. Given the approximate ansatz for the skyrmion profile (2) we consider that the agreement between theoretical and direct modelling calculations is excellent.
Particularly, the analytical model reproduce well the bi-stability of the skyrmion state at high fields.
In Fig. \ref{Energy_D} we present the
analytical skyrmion energy as a function of its diameter for PtCoPt dot near the critical field. The graph clearly demonstrates how an additional minimum corresponding to the large skyrmion diameter  appears near the dot edge, with the energy smaller than that for the skyrmion of the smallest diameter. The small size skyrmion becomes unstable at high values of the out-of-plane field. Note that a similar behavior was reported recently in Ref. [23] as a function of temperature. The skyrmion bi-stability at zero applied field was simulated recently in thick multilayer dots due to the effect of the dipolar interactions \cite{Zelent}. Importantly, our model also allows us to estimate the energy barrier between the states with the skyrmions of two diameters. They appears to be small, of the order of several $k_B T$  at room temparature meaning that at some values of the out-of-plane field the skyrmion will superparamagnetically fluctuate from one diameter to another. The shallow minima of the skyrmion energy vs. the skyrmion radius result in the low frequency breathing modes \cite{Mruczkiewicz2017}. The breathing mode frequency then is in sub-GHz range and can be close to the frequency of the lowest skyrmion gyrotropic mode related to its topological charge \cite{Guslienko2017}. When the out-of-plane magnetic field increases the skyrmion with large diameter become more and more stable against thermal fluctuations.
At the same time, the energy barriers between the small diameter skyrmion and the perpendicular saturated states (of the order of 30 - 40 $k_BT$) are large enough to provide the skyrmion thermal stability.

\begin{figure}[hbtp]
\includegraphics[width=7.5cm]{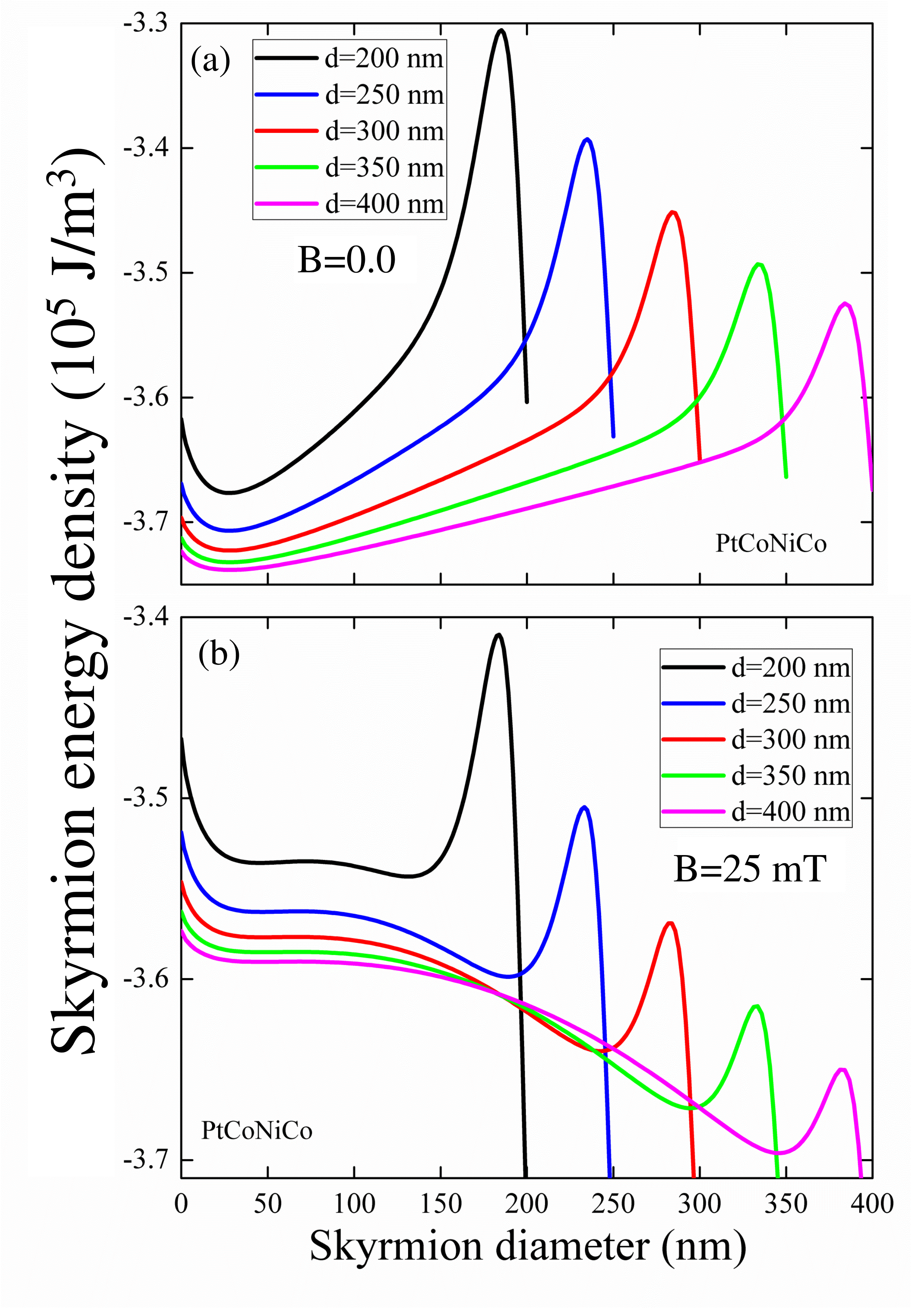}
\caption{Skyrmion energy density as a function of the skyrmion diameter in a dot of PtCoNiCo for different values of the dot diameter, calculated by analytical approach for the out-of-plane magnetic field (a) $B=0$ and (b) $B=25$ mT.}
\label{Energy_Dot}
\end{figure}

Another illustration is presented in Fig. \ref{Energy_Dot} for PtCoNiCo material and varying the dot diameter. We clearly observe that at zero field (Fig. \ref{Energy_Dot}(a)) the energy minimum is located near the small values of the skyrmion diameter and are only weakly displaced to the right for larger dot diameters. At the same time, at $B = 25$ mT (Fig. \ref{Energy_Dot}(b)) there are two minima: one is located at a small diameter value and the other one is located at large diameter value, stabilized by the dot boarder and with smaller magnetic energy. The minimum corresponding to the small skyrmion diameter is shallow and disappears increasing the bias field strength.

\begin{figure}[hbtp]
\centering
\includegraphics[width=7.5cm]{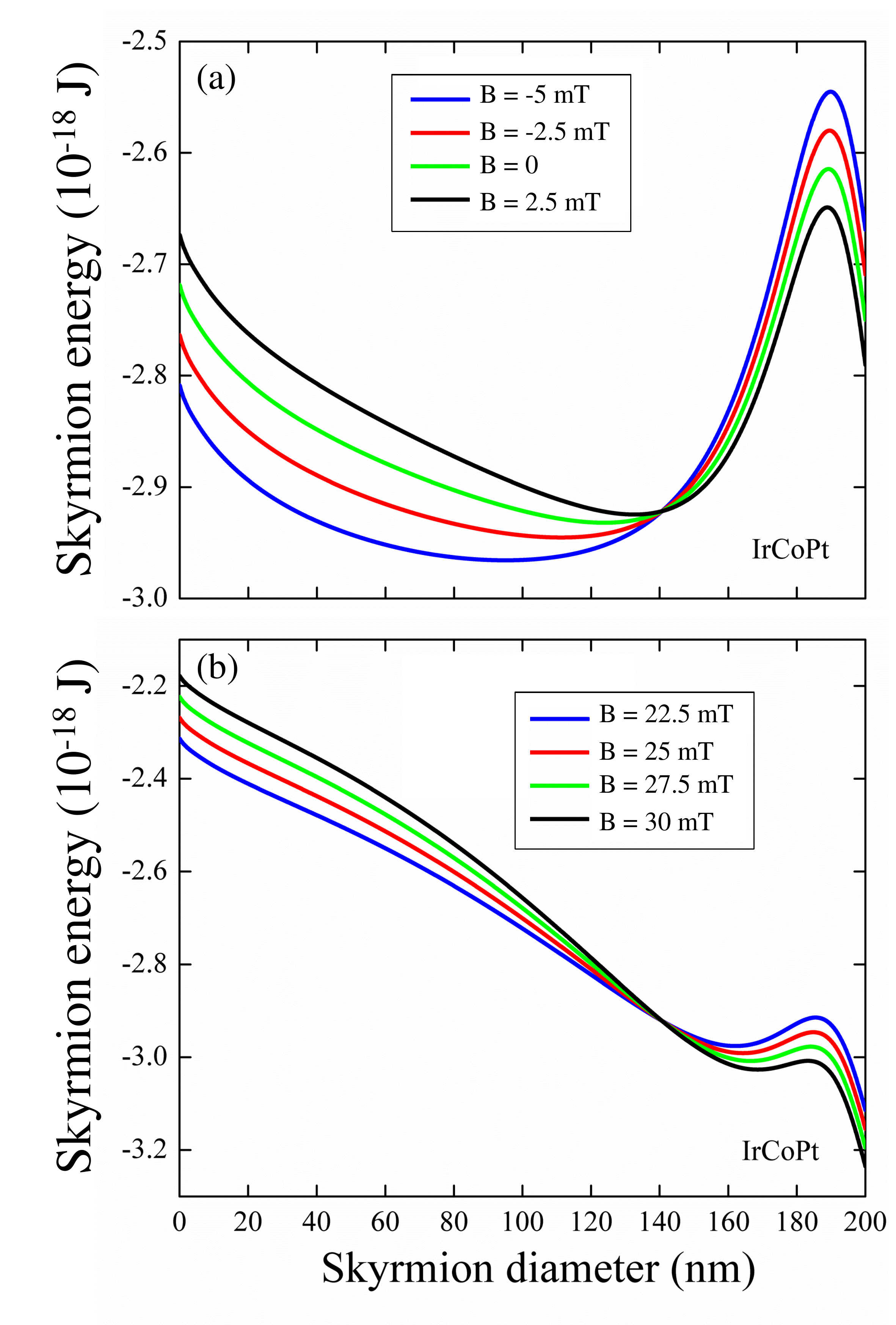}

\caption{Skyrmion energy in a IrCoPt dot of 200 nm in diameter, as function of the skyrmion diameter for different values of the out-of-plane magnetic field (a) in the stable region (b) in the metastable region.}
\label{Energy_D1}
\end{figure}

In the case of IrCoPt shown in Fig. \ref{Energy_D1}(a) in the skyrmion stability region, e.g., at zero out-of-plane magnetic field, the energy minimum is localized at large values of the skyrmion  diameters and is strongly displaced with a relatively small change in the bias field.
In the metastability region as in the example presented in Fig. \ref{Energy_D1}(b) this minimum is always localized near the dot edge and only weakly displaces to higher skyrmion diameter values with the field increase.

\section{Conclusions}
Micromagnetic simulations and analytical calculations were performed to study the size of the N\'eel skyrmion  confined in a cylindrical ultrathin nanodot. The results show a very good agreement between both methods allowing us to conclude about  good accuracy of the ansatz { given in Eq.(\ref{Anzats}) for the N\'eel skyrmion profile.

There are two different types of the skyrmions (large and small size) with distinct dependence of their radius on the applied out-of-plane field. The large radius skyrmions are rare, stable at zero field and possess strong and continuous dependence of their radius on the applied out-of-plane field and dot size. These skyrmions undergo the transition between stable and metastable states. In the metastable state their size dependence on the applied field becomes weak.

In the other, more frequent case, the skyrmions have small radius at zero field and are always metastable. They are characterized by a  weak dependence of their radius on the applied field until some critical field is reached at which the radius increases suddenly. The skyrmion energy is bi-stable close to this critical field and we observed the co-existence of a small and a large size metastable skyrmions within some field interval.


Therefore, we calculated a universal behavior of the Neel skyrmions in ultrathin dots with interface induced DMI. This opens a possibility to detect experimentally the skyrmion metastability/stability just looking at the skyrmion radius variation as a function of the applied out-of-plane magnetic field. We also calculated that the hysteresis process is a property of metastable skyrmions only. In this case, three  skyrmion states (all metastable) may co-exist: two states with the skyrmion cores are parallel to the field and one - antiparallel to it.\\

\section*{Acknowledgements}

F.T., A.R., and J.E. acknowledge financial support from the Fondecyt Grant No. 1150952,  DICYT Grant 041731EM-POSTDOC from VRIDEI-USACH and  Financiamiento Basal para Centros Cient\'{\i}ficos y Tecnol\'{o}gicos de Excelencia FB0807. CONICYT Ph.D. Program Fellowships is also acknowledged. K.G. acknowledges support by IKERBASQUE (the Basque Science Foundation).
The work of K.G. and O.C.-F. was supported by the Spanish Ministry of Economy and competitiveness under the project FIS2016–78591-C3-3-R.

\end{document}